\documentclass[preprint2,10.5pt]{aastex}
\usepackage{epsfig}
\begin{document}
\title{\large{\rm{THE GALACTIC CALIBRATION OF THE CEPHEID PERIOD-LUMINOSITY RELATION AND ITS IMPLICATIONS FOR THE UNIVERSAL DISTANCE SCALE}}}
\author{David G. Turner$^{1,2,6}$, Daniel J. Majaess$^{1,2}$, David J. Lane$^{1,2}$, Joanne M. Rosvick$^{3,6}$ \\ Arne A. Henden$^4$, David D. Balam$^5$}
\affil{$^1$ Saint Mary's University, Halifax, Nova Scotia, Canada}
\affil{$^2$ The Abbey Ridge Observatory, Stillwater Lake, Nova Scotia, Canada}
\affil{$^3$ Thompson Rivers University, Kamloops, British Columbia, Canada}
\affil{$^4$ AAVSO, Cambridge, Massachusetts, U.S.A.}
\affil{$^5$ Dominion Astrophysical Observatory, Victoria, British Columbia, Canada}
\affil{$^6$ Visiting Astronomer, Dominion Astrophysical Observatory, Herzberg Institute of Astrophysics, \\ National Research Council of Canada}
\email{\rm{turner@ap.smu.ca}}

\begin{abstract}
The Galactic calibration of the period-luminosity (PL) relation for classical Cepheids is examined using trigonometric, open cluster, and pulsation parallaxes, which help establish independent versions of the relationship. The calibration is important for the continued use of classical Cepheids in constraining cosmological models (by refining estimates for {\it H}$_0$), for defining zero-points for the SNe Ia and population II (Type II Cepheids/RR Lyrae variables) distance scales, for clarifying properties of the Milky Way's spiral structure, and for characterizing dust extinction affecting Cepheids in the Milky Way and other galaxies. Described is a program to extend and refine the Galactic Cepheid PL relation by obtaining {\it UBVRIJHK}$_{\rm s}$ photometry and spectra for stars in open clusters suspected of hosting classical Cepheids, using the the facilities of the OAMM, DAO, AAVSO, and ARO.
\end{abstract}

\section{Introduction}

  In the present era there is considerable interest in the distance scale established by classical Cepheid variables. The Cepheid period-luminosity relation is the primary standard candle used to establish the distances to galaxies hosting Type Ia supernovae, as well as to derive an accurate value for the Hubble constant $H_0$, which is necessary for constraining a variety of cosmological parameters, including the nature of dark energy. A considerable amount of effort has been spent in attempts to solidify the calibration and usefulness of the Cepheid distance scale, and yet the picture obtained from a perusal of the literature is that many questions about the zero-point and slope of the period-luminosity relation remain unanswered. Just how solidly established is the Cepheid distance scale?
  
\section{The Empirical Approach}

  Many researchers refer to this method as the ``theoretical'' approach, but a better terminology would be ``model dependent.'' A good start is the effective temperature-colour relation derived by Gray (1992) from a comparison of model atmosphere determinations of effective temperature {\it T}$_{\rm eff}$ in bright non-variable stars with their unreddened {\it B--V} colours:

\begin{eqnarray}
\log T_{\rm eff}=3.988-0.881(B-V)+2.142(B-V)^2 \\ \nonumber
-3.614(B-V)^3+3.2637(B-V)^4 \\ \nonumber
-1.4727(B-V)^5+0.2600(B-V)^6 \nonumber
\end{eqnarray}

\noindent The relationship can be used to infer mean effective temperatures for Cepheids from {\it B--V} colours corrected for reddening, even though the same colours in the case of variable stars are affected to a small degree by line blocking and line blanketing effects over the course of Cepheid pulsation cycles.

\begin{figure}[h]
\centerline{
\epsfig{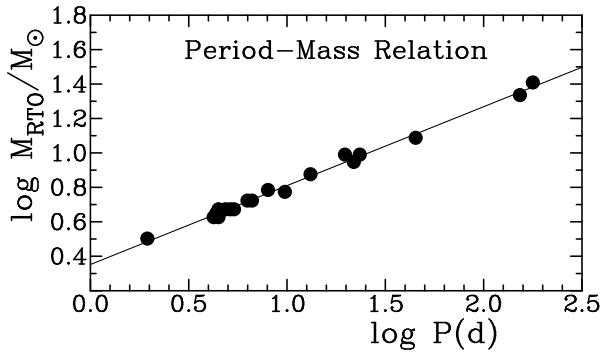}
}
\caption{\small{The period-mass relation for Cepheids and Cepheid-like supergiants in open clusters and associations. The slope of the relationship is $0.46\pm0.01$.}}
\label{fig1}
\end{figure}

  A sufficient number of Cepheids and Cepheid-like objects (V810 Cen, HD 18391) belonging to open clusters have been studied that one can obtain useful information on the Cepheids themselves from the empirical information gleaned for cluster stars. Such a study was made previously by Turner (1996), but has been updated for this study using results from more recent studies (Turner, Pedreros \& Walker 1998; Turner, Usenko \& Kovtyukh 2006; Turner et al.~2007, 2009). A period-mass relation can be inferred for cluster Cepheids, for example, by establishing the masses of cluster stars at the main-sequence red turn-off (RTO), marking the termination of core H-burning, as tabulated by Meynet, Mermilliod \& Maeder (1993) in their stellar evolutionary models. Results are presented in Fig.~\ref{fig1} for 19 Cepheids and Cepheid-like supergiants. In the original study (Turner 1996), the implied slope of the period-mass relation was $0.50\pm0.02$, implying a simple relationship of the type $M/M_{\odot} \sim P^{\frac{1}{2}}$.

  In Turner (1996), cluster ages were derived from matching the upper ends of the resulting cluster colour-magnitude diagrams to best-fitting model isochrones by Meynet et al.~(1993). For the present study, ages were inferred by the alternate technique of using the blue cluster turnoff points with the relations of Meynet et al.~(1993). The results (Fig.~\ref{fig1}) lead to a different slope for the relationship, namely $0.46\pm0.01$, slightly different from the Turner (1996) results, but close enough to confirm them. It appears that the analysis may need to be repeated with more up-to-date stellar evolutionary model results that generate identical cluster ages from cluster turnoff points and from isochrone fitting. The main point is that many Cepheid parameters may be related in simple fashion to pulsation period.

  A similarly straightforward parameterization applies to Cepheid radii, although that was not always the case. Cepheid radii can be established via the Baade-Wesselink (B-W) technique, in which phases of identical {\it T}$_{\rm eff}$ or surface brightness provide estimates for radius ratios at those phases via:

\begin{equation}
L_1/L_2=10^{-0.4(m_1- m_2)}=R_1/R_2 
\end{equation}

\noindent Since the differences {\it R$_1$--R$_2$} can be established between those phases through integration of a Cepheid's radial velocity changes, one can determine its mean radius using all phase pairs of identical {\it T}$_{\rm eff}$.

  Theory and practice differ, of course. For one thing, {\it B--V} colour does not correlate well with effective temperature. Alternate choices have included the indices {\it V--R}, {\it V--I}, and {\it V--K}, and the Brigham Young University {\it KHG} index (Turner, Leonard \& English 1987; Turner 1988). The {\it KHG} index, in particular, monitors atmospheric effective temperature in Cepheids using narrow-band filters tied to the Ca II K-line, H$\delta$, and the molecular G-band visible in Cepheid spectra, and has the advantage of being relatively independent of atmospheric and interstellar extinction. Use of the {\it KHG} index generates Cepheid radii (Turner 1988; Turner \& Burke 2002) in which the basic premise of the Baade-Wesselink method is satisfied, something not normally tested with more sophisticated approaches, for example those using Bayesian Markov-Chain Monte Carlo code. When the B-W method is done correctly, a plot of radius ratios versus radius differences should describe a tight clockwise loop (Turner 1988) and not an open counterclockwise loop. The test fails when {\it B--V} colour is used as the temperature indicator.

\begin{table}[h]
\caption{Slope of the period-radius relation.}
\label{tab1}
\vspace{2mm}
\centering
\begin{tabular}{cc}
\hline \hline
\noalign{\smallskip}
Slope &Source \\
\noalign{\smallskip}
\hline
\noalign{\smallskip}
0.70 &Cogan (1978) theory \\
0.70: &Gieren (1981) \\
0.587--0.956 &Fernie (1984) optimum 0.824 \\
0.84 &Gieren (1984) \\
0.63 &Coulson, Caldwell \& Gieren (1986) \\
0.77 &Gieren (1986) \\
0.743 &Gieren, Barnes \& Moffett (1989) \\
0.750 &Gieren, Fouqu\'{e} \& G\'{o}mez (1998) \\
0.751 &Laney \& Stobey (1995) \\
0.747 &Turner \& Burke (2002) \\
\noalign{\smallskip}
\hline
\end{tabular}
\end{table}
     
\begin{figure}[h]
\centerline{
\epsfig{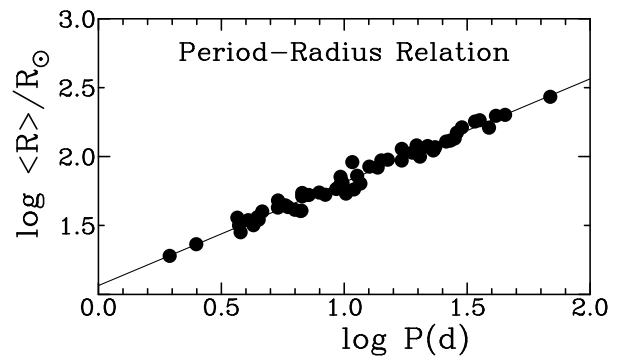}
}
\caption{\small{The period-radius relation delineated by data from Laney \& Stobey (1995) and Turner \& Burke (2002). The slope of the relationship is 0.75.}}
\label{fig2}
\end{figure}

  Results of past studies establishing the slope of the Cepheid period-radius relation are presented in Table~\ref{tab1}. The derived slope varied widely from study to study in the early years, but eventually converged upon a value of $0.750\pm0.003$ a decade ago, as indicated by the results of Gieren et al.~(1998), Laney \& Stobey (1995), and Turner \& Burke (2002). Data for the latter two provide a tightly defined period-radius relation, as indicated for the last two sources by the data of Fig.~\ref{fig2}. The implication is that $\langle R \rangle/R_{\odot} \sim P^{\frac{3}{4}}$.

  Not all results can be combined, however. That is because it is necessary to correct the measured radial velocity variations in Cepheids for projection effects arising from general envelope pulsation in the stars. The corresponding projection factor {\it p} is close to 1.30, but has also varied significantly over the past 25 years. Since {\it p} scales the resulting radii, one source of Cepheid radii may scale differently from another because of differences in adopted {\it p} values. That results in a zero-point shift but not a change in slope. The manner in which the choice of {\it p} has varied over the years is indicated in Table~\ref{tab2}. The abrupt increase in the parameter occurring in the mid 1980s has recently been reversed, and perhaps agreement will eventually settle upon the original values near 1.30--1.31. In fact, the exact value can be tested in simple fashion, as noted below.

\begin{table}[h]
\caption{Estimates of the B-W projection factor {\it p}.}
\label{tab2}
\vspace{2mm}
\centering
\begin{tabular}{cc}
\hline \hline
\noalign{\smallskip}
{\it p} &Source \\
\noalign{\smallskip}
\hline
\noalign{\smallskip}
1.412 &Getting (1935) \\
1.31 &Parsons (1972) \\
1.31 &Karp (1975) \\
1.31--1.47 &Hindsley \& Bell (1986) \\
1.34--1.38 &Gieren et al. (1989) \\
1.30--1.42 &Gray \& Stevenson (2007) \\
1.19--1.31 &Laney \& Joner (2009) \\
1.30--1.31 &Region of overlap (all studies) \\
\noalign{\smallskip}
\hline
\end{tabular}
\end{table}

\begin{figure}[h]
\centerline{
\epsfig{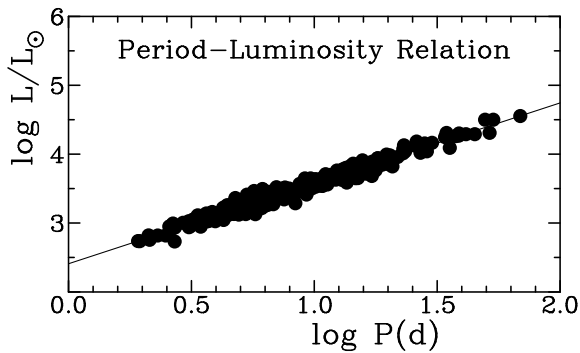}
}
\caption{\small{The period-luminosity relation defined by Cepheids of well-established reddening.}}
\label{fig3}
\end{figure}

  The Cepheid period-luminosity relation can therefore be constructed from first principles without regard to observational data, provided one has a reliable bolometric correction scale to convert mean absolute visual magnitudes $\langle M_V \rangle$ to luminosities in solar units $L/L_{\odot}$. Such a technique was adopted by Turner \& Burke (2002) and Turner (2010). An example is shown in Fig.~\ref{fig3} for stars of well-established reddening, where the colour excesses originate in studies such as those of Turner (2001), Laney \& Caldwell (2007), and Kovtyukh et al.~(2008), often with overlap between studies, many of which are based upon earlier studies of space reddening and spectroscopic reddening for Cepheids.

\section{The Observational Approach}

\begin{figure}[!t]
\centerline{
\epsfig{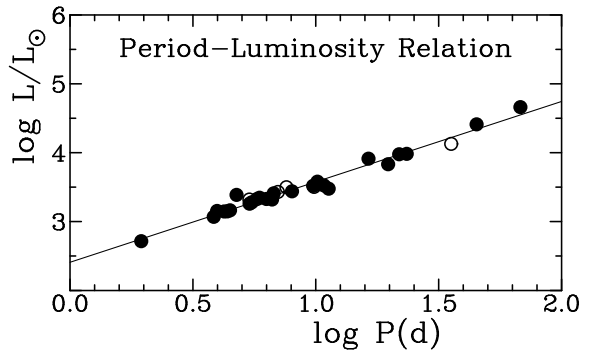}
}
\caption{\small{The period-luminosity relation defined by cluster (filled circles) and HST parallax (open circles) Cepheids.}}
\label{fig4}
\end{figure}

  The observational approach to the problem has previously used observations for Cepheids in the Large Magellanic Cloud (LMC) to define the period-luminosity relation, with a few Galactic calibrators to tie down the zero-point. In many cases the results for cluster Cepheids are artificially ``adjusted'' to account for changes to the Hyades/Pleiades zero-point for zero-age main sequence (ZAMS) fitting or to increase the stated precision of the results. But that is no longer necessary. The Hubble Space Telescope (HST) program of Benedict et al. (2007) to derive parallaxes for 10 relatively nearby Cepheids and the study of Turner (2010) for 24 Cepheids in Galactic clusters provide by themselves a reasonably large sample of calibrators for the period-luminosity relation. The results, presented in Fig.~\ref{fig4}, fit the identical relationship derived in Fig.~\ref{fig2} for Cepheids of well-established reddening:

\begin{equation}
\log L/L_{\odot}=2.409+1.168\log P 
\end{equation}

  Parallaxes from the Hipparcos catalogue prove to be less useful for such purposes (see Turner 2010), since there are peculiarities with the stated uncertainties, and in the parallaxes themselves, that create problems with the identification of proper pulsation mode (fundamental mode, overtone) for individual Cepheids. Nevertheless, a few of the most reliable Hipparcos parallaxes appear to confirm the scale of Cepheid luminosities derived from cluster and HST parallaxes. A series of tests is presented by Turner (2010), the conclusion being that an observational approach to calibrating the period-luminosity relation using Cepheids in Galactic clusters and Cepheids with HST parallaxes strongly confirms the relationship inferred by empirical means, namely use of a {\it T}$_{\rm eff}-(B-V)$ calibration, a well-defined period-radius relation, and a calibrated scale of bolometric corrections (Turner \& Burke 2002). Incidentally, that conclusion in itself appears to confirm the choice of a projection factor of $p=1.30-1.31$ for Cepheid B-W studies, as noted earlier. Galactic calibrators also span a wide range of pulsation periods ranging from $2^{\rm d}$ to $68^{\rm d}$, making them ideal for calibration purposes, although a few cluster Cepheids fit the relationship somewhat poorly, even with colour spread in the instability strip taken into account (Turner 2010). That situation may improve with further study of the associated star clusters.

\begin{figure}[h]
\centerline{
\epsfig{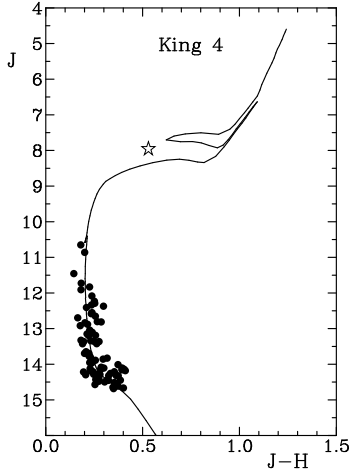}
}
\caption{\small{A \textit{preliminary} 2MASS colour-magnitude diagram for the cluster King 4, with the Cepheid UY Per indicated by a star symbol. A 10$^8$-yr Padova isochrone is shown for a reddening of {\it  E(J--H)} = 0.25 ({\it E(B--V)} = 0.89) and {\it J--M}$_J \gtrsim 12.4$ ({\it V}$_0${\it--M}$_V$ = 11.75, $d \gtrsim 2.24$ kpc).  Deeper $JHK_s$ and $VI$ colour-magnitude diagrams constructed from OAMM and ARO data are forthcoming.}}
\label{fig5}
\end{figure}

  A program has been initiated to increase the number of Galactic calibrators through the study of relatively unstudied open clusters that are spatially coincident, or nearly spatially coincident, with well-studied Cepheid variables. The program is rather ambitious and observations are being obtained from the Dominion Astrophysical Observatory (DAO), l'Observatoire Astronomique du Mont M\'{e}gantic (OAMM, Artigau et al.~2009, 2010), the Abbey Ridge Observatory (ARO), and the Sonoita Research Observatory of the American Association of Variable Star Observers (AAVSO). Surprisingly, there are several good cases of cluster Cepheids that have yet to be studied extensively. The case for UY Per as an outlying member of the open cluster King 4  (Turner 1977) is shown as an example in Fig.~\ref{fig5}.

\begin{figure}[h]
\centerline{
\epsfig{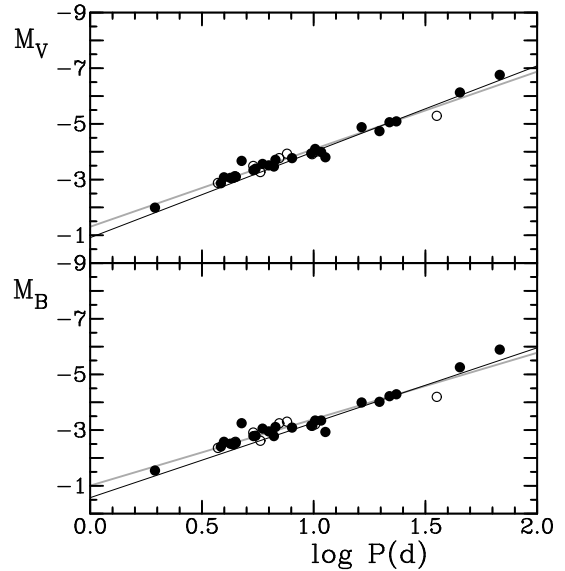}
}
\caption{\small{Absolute magnitudes $M_V$ and $M_B$ for Cepheids in open clusters (filled circles) and with HST parallaxes (open circles) produce the gray lines as best-fitting relationships, compared with the black lines predicted by the Sandage, Tammann \& Reindl (2004) calibration.}}
\label{fig6}
\end{figure}

  A ramification of using the scale of Cepheid luminosities cited here and by Turner (2010) is that it affects distances derived for distant galaxies, for example from the distance scale of Sandage, Tammann \& Reindl~(2004). A comparison of Cepheid absolute magnitudes $M_V$ and $M_B$ derived from cluster and HST parallaxes with respect to the Sandage et al.~(2004) Galactic calibration is shown in Fig.~\ref{fig6}. Here the scatter is related to the location of each Cepheid in the instability strip. The best-fitting relationships for cluster and HST data from least squares and non-parametric fits are:

\begin{eqnarray}
\nonumber
\langle M_V \rangle=-1.304\pm0.065 -2.786\pm0.075\log P \\ 
\langle M_B \rangle=-1.007\pm0.087 -2.386\pm0.098\log P 
\end{eqnarray}

  By comparison, the Sandage et al.~(2004) relationships underestimate the luminosities of Cepheids with periods less than $20^{\rm d}$ and overestimate the luminosities of Cepheids with periods in excess of $20^{\rm d}$. The latter, of course, are the objects most likely to be used to establish distances to galaxies hosting Type Ia supernovae. The effect appears small in Fig.~\ref{fig6}, but is more pronounced using a Wesenheit, or reddening-free, formulation, and can affect derived distances to galaxies by 10\% or more (Majaess 2010b). The effects on the Type Ia supernova calibration of Sandage et al.~(2006) may be significant. For example, they can account for the difference between the value of $H_0=62.3\pm5.0$ km s$^{-1}$ Mpc$^{-1}$ derived by Sandage et al.~(2006) for the Hubble constant applicable to galaxies hosting type Ia supernovae and the comparable value of $H_0=71\pm6$ km s$^{-1}$ Mpc$^{-1}$ derived by Freedman et al.~(2001) (Benedict et al.~2007, Majaess 2010b).  Reducing the uncertainty associated with the Hubble constant to $\le 5\%$ is a primary objective going forward (e.g, Ngeow et al.~2009; Macri \& Riess 2009).

  Another complication is that Cepheids at the limits of detectability in distant galaxies are frequently located in crowded fields where it can be difficult to extract uncontaminated light curves for the variables. Photometric errors can generate deleterious effects that may bias the determination of accurate period-absolute magnitude relations for Cepheids in such galaxies, thereby affecting their use as distance indicators (e.g., Mochejska et al.~2004). An example is provided by Turner (2010) for the galaxy NGC 4258. Wesenheit magnitudes for Galactic Cepheid calibrators exhibit a scatter of at most $\pm0^{\rm m}.5$ as a function of period, compared with a scatter in excess of $\pm1^{\rm m}.0$ for the Wesenheit magnitudes of Cepheids in the crowded inner regions of NGC 4258.  The situation may worsen for more distant galaxies used to calibrate the distance scale for type Ia supernovae.
  
\section{Type II Cepheids and RR Lyrae Variables}
An additional complication for the classical Cepheid distance scale is the importance of metallicity to the zero-point of the period-luminosity relationship. An advantage gained from a Galactic calibration is that it applies to Cepheids of roughly solar metallicity, much like the expected metallicity of Cepheids in the sample of spiral galaxies used to calibrate the Type Ia supernova relation. Tests of the importance of metallicity have traditionally been done by comparing the luminosities of Cepheids in the metal-rich inner regions of galaxies with those in their comparably metal-poor outer regions. That makes the possibility of contamination by crowding an important consideration (Mochejska et al.~2004; Majaess, Turner \& Lane 2009c; Majaess 2010b).

\begin{figure}[h]
\centerline{
\epsfig{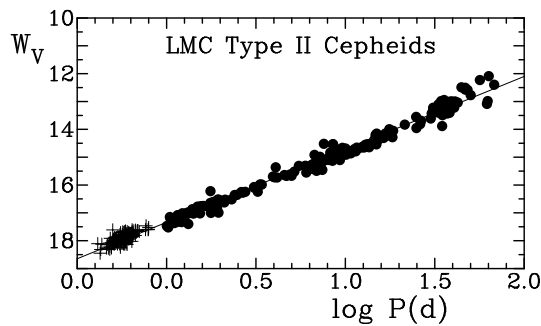}
}
\caption{\small{Mean brightnesses from OGLE data of LMC Type II Cepheids (filled circles: BL Her, W Vir, RV Tau variables) and RR Lyrae variables (crosses) as a function of the fundamentalized pulsation period.  The observations are from Soszy{\'n}ski et al.~(2008, 2009).}}
\label{fig7}
\end{figure}

  An alternate approach can be made using Type II Cepheids and RR Lyrae variables, which also appear to exhibit a tight relationship between luminosity and pulsation period, thereby permitting a comparison with galaxy distance moduli obtained from classical Cepheids.  That comparison implies that {\it VI}-based reddening-free Cepheid relations are comparitively insensitive to metallicity (Majaess, Turner \& Lane 2009c; Majaess 2009, 2010b, see also Udalski et al.~2001 and Pietrzy{\'n}ski et al.~2004).  Reddening-free Wesenheit magnitudes for Type II Cepheids and RR Lyrae variables in the LMC are plotted in Fig.~\ref{fig7}. The mean Wesenheit magnitudes for such stars appear to follow a linear relationship that links the luminosities of RR Lyrae variables and Type II Cepheids (Matsunaga et al.~2006; Majaess 2009), although the linearity seems to break down for the long pulsation periods corresponding to RV Tauri variables.

\begin{figure}[!t]
\centerline{
\epsfig{file=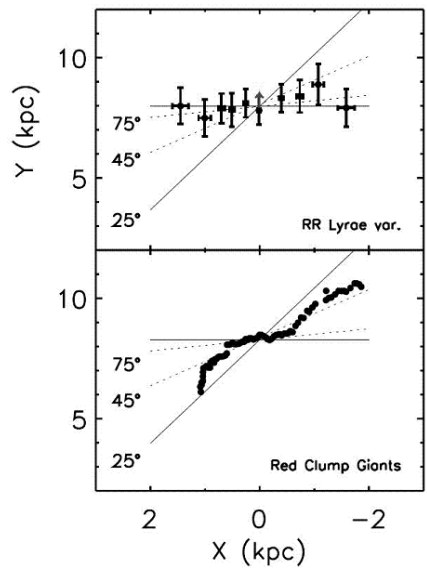, scale=0.35}
}
\caption{\small{Mean distances from OGLE data for RR Lyrae variables in the Galactic bulge (upper) relative to peaks in the distribution of red clump stars (lower) mapped by Nishiyama et al.~(2005).}}
\label{fig8}
\end{figure}

  Type II Cepheids also have the potential for use in other studies. Classical Cepheids, for example, have long been used to map the nearby spiral arms of the Galaxy, the most recent studies being those of Majaess, Turner \& Lane (2009a,b) and Turner \& Majaess (2010). Their older Type II and RR Lyrae counterparts are more suitable for studying other characteristics of the Galaxy, for example by providing an independent estimate for the distance to the Galactic centre (Majaess, Turner \& Lane 2009a, Majaess 2010a), or evidence for a Galactic bar. Fig.~\ref{fig8} is a plot of the mean distances of RR Lyrae variables detected in directions towards the Galactic centre, based upon {\it VI} photometry for RR Lyrae variables in the direction of the Galactic bulge from Collinge, Sumi \& Fabrycky (2006). A comparison with the bar-like structure at the Galactic centre mapped by Nishiyama et al.~(2005) using red clump stars reveals an apparent discrepancy, the RR Lyrae variables do not delineate a prominent bar, instead being concentrated concentrically within the Galactic bulge. Conceivably the bar of Nishiyama et al.~(2005) formed much later than the mean epoch of formation for the precursors of the RR Lyrae variables, or possibly it is a simple difference in population types (Alcock et al.~1998), given that RR Lyrae variables are a common constituent of the Galactic halo.

\section{Summary}

  As noted here, accurate period-luminosity or period-absolute magnitude relations for classical Cepheids can be constructed using Galactic calibrators tied to the data of Benedict et al.~(2007) and Turner (2010). The resulting linear relations closely match independent relationships derived using the Cepheid period-radius relation, a scale of Cepheid effective temperatures inferred from unreddened mean $(\langle B \rangle-\langle V \rangle)_0$ colours, and a reliable scale of bolometric corrections. Galactic calibrators also have the advantage of being bright and relatively nearby, whereas their counterparts in the Large and Small Magellanic Clouds are considerably fainter. The latter are used with Type II and RR Lyrae calibrators to test the importance of metallicity for the extragalactic distance scale.

\end{document}